\documentclass[12pt,cites,graphicx]{article}  
\usepackage{epsfig,rotating}  
  
\title{A new interpretation of the dynamic structure model of ion transport  
in molten and solid glasses}  
 
\author{Armin Bunde,$^a$ Malcolm D. Ingram$^b$ and Stefanie Russ$^c$\\ 
$^a$ Institut f\"ur Theoretische Physik III, Justus-Liebig-Universit\"at Giessen,\\ 
D-3592 Giessen, Germany. E-mail: Armin.Bunde@physik.uni-giessen.de \\ 
$^b$ Department of Chemistry, University of Aberdeen, Aberdeen AB24 3UE, \\ 
Scotland, UK. E-mail: m.d.ingram@abdn.ac.uk \\
$^c$ Institut f\"ur Theoretische Physik III, Justus-Liebig-Universit\"at Giessen,\\ 
D-3592 Giessen, Germany. E-mail: Stefanie.Russ@physik.uni-giessen.de \\ } 
\begin{document}  
\maketitle  
\renewcommand{\thefootnote}{\fnsymbol{footnote}}

\noindent We explore progress in understanding the behaviour of cation conducting glasses,  
within the context of an evolving ''dynamic structure model'' (DSM). This
behaviour includes: in single cation glasses a strong dependence of ion
mobility on concentration, and in mixed cation glasses a range of anomalies
known collectively as the mixed alkali effect. We argue that this rich
phenomenology arises from the emergence during cooling of a well-defined
structure in glass melts resulting from the interplay of chemical
interactions and thermally driven ionic motions.
The new DSM proposes the existence of a new site relaxation process,
involving the shrinkage of empty $\bar A$ sites (thus tailored to the needs
of $A^+$ ions), and the concurrent emergence of empty $C'$'sites, which
interrupt the conduction pathways. This reduction of $\bar A$ sites is
responsible in the molten glass for the sharp fall in conductivity as
temperature drops towards $T_g$. The $C'$ sites play an
important role also in the mixed alkali effect, especially in regard to the
pronounced asymmetries in diffusion behaviour of dissimilar cations.
 
\section{Introduction}  
\label{intro}  
Since 1990, there has been growing interest in the mechanisms of ion
transport in disordered materials including glasses. This interest has been
sustained not only by a rich diversity of behaviour
\cite{ingram87,ngai96,funke02}, but also by rapid advances in spectroscopy,
notably in magic angle spinning NMR and in EXAFS \cite{ratai02,greaves90},
and in the increasing sophistication of molecular dynamics (MD) and other
computational techniques \cite{habasaki97,cormack02,sunyer02,heuer02}.
 
Much effort has gone into accounting for ''universal'' aspects of glass
behaviour which include the scaling properties of conductivity spectra
\cite{sidebottom96,sidebottom01}. Thus, through the work of Dyre and of
Funke et al \cite{funke02,schroder02}, conductivity spectra for disordered
materials are well described, even though there is still active discussion
of the underlying mechanisms.
 
The important challenge, which needs to be addressed, lies in creating a
conceptual framework to enable the behaviour of different glasses to be
compared and if possible, predicted from first principles. Some years ago, a
start was made in this direction by Bunde, Ingram and Maass through
introduction of the dynamic structure model (DSM). The essential idea
\cite{bunde91,maass92,bunde94,maassdr}, which was based originally on EXAFS
data of Greaves et al \cite{greaves90}, is that the mobile ions play a
decisive role in creating the sites which they occupy (as well as the
doorways to these sites) within the glassy matrix. In essence, the mobile
ions shape their immediate environments to meet their own requirements and
also leave behind empty sites (called $\bar A$ for $A^+$ cations, $\bar B$
for $B^+$-cations) which act as ''stepping stones'' and thus define the
conduction pathways. Any empty sites away from the conduction pathways, or
where any memories of previous cations have been forgotten, are referred to
as $\bar C$ sites.
 
The growth of such conduction pathways \cite{bunde94} explained the steep
rise in conductivity in single ionic glasses which accompanies increases in
mobile ion concentration, $c$, in many systems as a percolation phenomenon.
Thus, an effective power-law behaviour often applies, where the conductivity
scales as
\begin{equation} 
\sigma\sim c^{\alpha/kT}. 
\end{equation} 
$T$ is the temperature, $k$ is the Boltzmann constant and $\alpha$ is an
empirically determined parameter with the units of energy. This expression
was deduced first from Monte Carlo (MC) simulations and then confirmed from
an extensive survey of published experimental data \cite{maassdr}. A similar
steep increase in conductivity with ionic content also emerged from the
calculations based on the counter-ion model \cite{knoedler96} and on energy
landscape models (\cite{hunt97}, see also \cite{hunt94}).
 
In mixed cation glasses, discussion in terms of the DSM drew attention to
the consequences of the strong competition between dissimilar mobile ions in
their attempts to establish their conduction pathways. MC simulations showed
how this competition led to the fragmentation of pathways, and hence to the
sharp falls in conductivity that are among familiar characteristics of the
mixed alkali effect (MAE). The underlying cause was attributed to dissimilar
ions ($A^+$ and $B^+$) being unwilling to visit each other's ($\bar B$ and
$\bar A$) sites, because of a mismatch between the requirements of the ion
with what the ''wrong'' target site and the doorway to it could offer in
terms of cavity size and perhaps numbers of nearest neighbours.
 
The DSM continues to influence research in glassy ionics. Thus, although
there are some researchers, see for example \cite{kirchheim02}, who discuss
the origin of mixed cation effects without invoking site recognition or
percolation effects, there is a growing consensus that this is the right
approach. Thus, evidence for the existence of distinguishable $\bar A$ and
$\bar B$ sites comes from both MD and molecular orbital simulations
\cite{greaves90,Balasubramanian93,Uchino92,smith97}. Recently, Swenson et al
\cite{swenson03} have shown by reverse Monte Carlo (RMC) and bond-valence
calculations how $Li^+$ and $Rb^+$ ions in mixed-cation phosphate glasses
find themselves on pathways containing mixtures of $\bar{Li}$ and $\bar{Rb}$
sites. They too attribute the loss of ion mobility to a mismatch energy
along the lines previously proposed by some of us \cite{bunde94}.
 
As a consequence of this activity, new questions continue to arise. First
and foremost, there is the role of the network. How much does the matrix
participate in inter-site conversions ($\bar A$ to $\bar C$, etc.)? Do these
processes occur above and below $T_g$, or only above $T_g$ as suggested
recently by Maass \cite{maass99}? How closely linked are the chemical and
physical processes which lie behind the behaviours observed in molten and
glassy materials? Are we yet in a position to say that we have a theory of
the mixed alkali effect? The present paper is an attempt to see how far
these questions can be answered in the light of an updated dynamic structure
model.

\section{Single cation systems} 
\label{single}  
\subsection{Different temperature regimes}   
\label{diff_temp}  
Figure~\ref{bi:arrhenius} contains Arrhenius plots showing how the
conductivity varies exponentially with reciprocal temperature for
$Na_2O-2SiO_2$, $K_2O-2SiO_2$ and $0.7AgI-0.3Ag_2MoO_4$ glasses in both the
molten and glassy states \cite{souquet94}. Characteristically, the plots are
curved above $T_g$ and linear below. For convenience, we identify three
regions of interest. First, there is the high temperature regime, where
conductivities approach those of typical molten salts
(ca.~$1\,\Omega^{-1}cm^{-1}$). Second, there is a region of under cooling
which extends right down to $T_g$. This region is associated with huge
increases in melt viscosity (from about $10^1$ to $10^{13}\,Pa.s$) and by
comparison a rather smaller decrease (only two to three orders of
magnitude), in the ionic conductivity. Third, there is the region below
$T_g$ in which ion transport proceeds with a constant activation within an
apparently rigid solid. The conductivity decreases comparatively rapidly
with $1/T$ in the middle region, while it decreases comparably slower in the
first and third regimes.
 
\unitlength 1.85mm 
\vspace*{0mm} 
{ 
\begin{figure} 
\begin{picture}(80,50) 
\def\epsfsize#1#2{0.3#1} 
\put(3,-2){\epsfbox{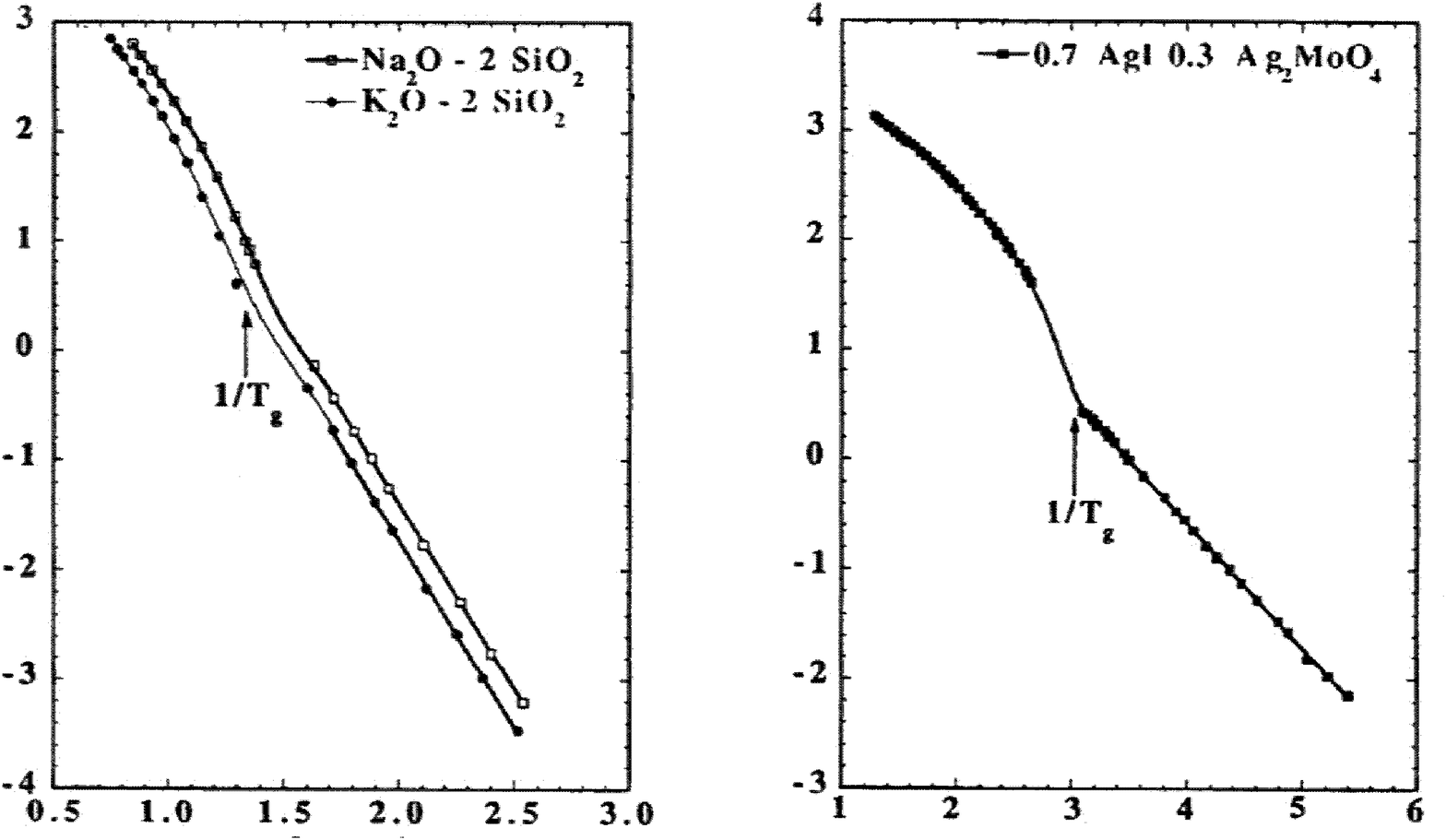}} 
\put(12,7){\makebox(1,1){\bf\large (a)}}  
\put(58,7){\makebox(1,1){\bf\large (b)}} 
\put(30,0){\makebox(1,1){\bf $10^3/T \, [K^{-1}]$}}  
\put(75,0){\makebox(1,1){\bf $10^3/T \, [K^{-1}]$}}  
\put(0,20){\begin{sideways} 
          \bf $\log (\sigma T) \, [\Omega^{-1} cm^{-1} K]$ 
          \end{sideways} 
         } 
\put(46,20){\begin{sideways} 
          \bf $\log (\sigma T) \, [\Omega^{-1} cm^{-1} K]$ 
          \end{sideways}  
          }                                                                  
\end{picture} 
\caption[]{\small Ionic conductivity as a function of temperature for (a) $Na_2O-2SiO_2$
  and $K_2O-2SiO_2$ and (b) $0.7AgI-0.3Ag_2MoO_4$ glasses in both the molten
  and the glassy state. (Redrawn from Ref.~\cite{souquet94}).  }
\label{bi:arrhenius} 
\end{figure}} 
 
At temperatures far above $T_g$, (region 1), all ionic motions are coupled
to processes involved in viscous flow. These microscopic processes include
the switching of non-bridging oxygens (nBOs) between neighbouring silicon
atoms and temporary changes in the coordination number of $Si$ from $4$ to
$5$ or even to $6$ \cite{stebbins91}. The sites created by the mobile ions
may be of limited duration, but they will be precisely determined by the
interplay of forces representing the chemical requirements of both mobile
ions and the various silicate or other anionic species present in the melt.
 
However, it is known from the work of Moynihan and Angell
\cite{howell74,angell86} that undercooling leads to progressive decoupling
of cation motions from the host matrix, with an associated increase in the
ratio of relaxation times for viscous flow to those for conductivity, i.e.
of $\tau_{\rm{shear}}/\tau_{\rm{sigma}}$. The value of this ratio at $T_g$
is called the decoupling index, $R_\tau$, and reaches values around
$10^{11}$ in typical silicate glasses. In practice, this means that since
structural relaxation times are approximately $100\,s$ at $T_g$, (as
measured, for example, in a typical differential scanning calorimetry (DSC)
experiment), the corresponding electrical relaxation time ($\tau = RC$) is
about $10^{-9}\,s$. In simple terms, this gives an indication of how long
ions are remaining in their sites at the glass transition temperature. The
inverse of $\tau$ is the average effective hopping rate.
 
On further cooling below $T_g$, the structure changes much less, but ions
spend increasingly longer times in their sites. At room temperature, a
sodium ion in a typical silicate glass might sit for approximately $1\,s$ in
its site before moving on. The corresponding relaxation time for viscous
flow is so long it would better be expressed on geological time scales
\cite{zanotto99}.
 
In the present context, it is noteworthy that Binder et al \cite{binder99}
also draw attention to the second temperature region described above, which
they identify as falling between the critical temperature $T_c$ of mode
coupling theory (MCT) where the system undergoes some ''structural arrest''
and the ''solidification temperature'' (i.e. the experimental $T_g$). They
also point out that the time scale for viscous flow at $T_c$ is about ten
orders of magnitude shorter than it is at $T_g$, and suggest that it is
unlikely that significant changes will not occur as a liquid is cooled
between $T_c$ and $T_g$. This is precisely the region of melt behaviour
where the decoupling ratio is increasing from values close to unity to much
larger values. Indeed, within the framework of MCT it seems likely that near
$T_c$ the dynamics of the network forming species will be dominated by
mutual blocking effects, and their diffusion will be drastically reduced. In
effect, the network structure is forming around $T_c$, and already a pattern
of behaviour is being established which continues on cooling even into the
solid glass.
 
\subsection{The mechanism of ion transport} 
\label{mech} 
 
Evidence from recent NMR studies \cite{jones01} shows that cations in glass
will typically find sites near at least one nBO. More generally, in glasses
where nBOs may be absent, e.g. in aluminosilicate glasses, the cations will
be found in regions close to where negative charge resides in the network.
The increasing values of $\tau_{\rm{shear}}/\tau_{\rm{sigma}}$, see above,
show that even in the melt the local network structure persists for much
longer times than the ions reside in their sites. We are looking at a
hopping process which is well established in the molten state.
 
But what kinds of site do the ions hop into? What distinguishes an $\bar A$
from a $\bar C$ site has not so far been rigorously defined. It was assumed
\cite{bunde94} that $\bar A$ character expresses a ''goodness of fit'', and
combines a number of factors that determine the accessibility of the site to
an $A^+$ ion, including cavity size, coordination number, and the
availability of negative charge as indicated often by the presence of nBOs.
 
We now propose to distinguish two kinds of $\bar C$ sites. First, there are
$\bar C$ sites, which in silicate melts would be remote from nBOs.  These
can only become $\bar A$ sites with the active assistance of chemical
rearrangements occurring in the silicate matrix. It is likely that such
rearrangements occur readily only at high temperatures (possibly in the
vicinity of the $T_c$ of MCT), and that they occur less and less frequently
as the melt is cooled towards $T_g$.
 
The second kind of $\bar C$ site has very different origins. Let us suppose
that a freshly vacated $\bar A$ site looks just like a filled site, but with
the cation missing.  As long as it stays like that, we shall call it an
$\bar A$ site. However, after being vacated, the $\bar A$ site may start to
relax in such a way as to minimise the (now largely uncompensated)
repulsions between negatively charged oxygens in what was previously the
primary coordination sphere of the cation. This relaxation might be
envisaged, for the purpose of illustration, as the reorientation of $Si-O$
bonds to avoid pointing the oxygens directly at each other (as will be the
case in the newly emptied $\bar A$ site) but instead towards neighbouring
silicon atoms which (because of differences in chemical electronegativity)
will carry small positive charges. Whatever the actual (microscopic) details
of this relaxation process, we call this new site a $C'$ site.
 
The processes of site relaxation occur in the molten state, but because
melts are relatively rich in ''free volume'', the $C'$ and $\bar A$ sites
may not differ too much from each other.
However, the melt contracts on cooling and so some free volume must
disappear. We suggest that this contraction occurs most readily by the
conversion of empty $\bar A$ into smaller $C'$ sites, as illustrated in
Fig.~\ref{bi:sketch}. Below $T_g$, the total number of $\bar A$ and $C'$
sites will remain constant and further shrinkage of the $C'$ sites will be
less important.
 
\unitlength 1.85mm 
\vspace*{0mm} 
{ 
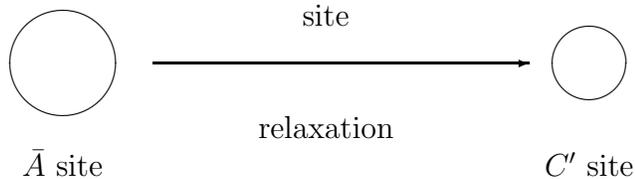
\begin{figure} 
\begin{picture}(80,20) 
\def\epsfsize#1#2{0.4#1} 
\put(0,0){\unitlength1mm 
\linethickness{0.4pt}  
\put(52,20){\vector(1,0){50}} 
 
\put(40,20){\circle{20}}  \put(110,20){\circle{10}} 
 
\put(70,25){\makebox(10,10)[b]{site}} 
\put(65,10){\makebox(20,10)[b]{relaxation}} 
 
\put(30,5){\makebox(20,10)[b]{$\bar A$ site}} 
\put(100,5){\makebox(20,10)[b]{$C'$ site}} 
} 
\end{picture} 
\caption[]{\small Site relaxation, showing 
  the collapse of an empty $\bar A$ site, either during cooling of molten
  glass or when glass is subjected to external pressure, with the consequent
  creation of a new $C'$ site and the resulting disappearance of some
  free volume at the same time.  }
\label{bi:sketch} 
\end{figure}} 
 
Below $T_g$, two alternative scenarios may in fact be envisaged. In the
first, the structure may be completely frozen, with all sites retaining
their $\bar A$, $C'$, and $\bar C$ identities as they were defined at $T_g$.
Such a situation would be consistent with many contemporary views of glass
behaviour \cite{knoedler96,maass99}.
 
The alternative would be that minor structural fluctuations such as those
involved in converting $C'$ into $\bar A$ sites or vice versa, would still
occur, since changes in network topology would not be involved. The main
difference between $C'$ and $\bar A$ sites would thus be related to local
density fluctuations. In this way, it would still be possible to refer to a
dynamic energy landscape, which is attributed to the network structure, even
below $T_g$. 

Whichever scenario is adopted, this new description has several advantages.
As one example, it allows one to visualise what is happening in the
Arrhenius plots (see Fig.~\ref{bi:arrhenius}) in different temperature
regimes. Above $T_g$, the progressive increase in slope seen during cooling,
is a consequence of the decreasing number of empty $\bar A$ sites as they
are converted into smaller $C'$ sites. The arrest at $T_g$ signals the
effective end of this process. Below $T_g$, the energy landscape
is determined by the distribution of empty $\bar A$ and $C'$ sites.

We can go on further to consider that the number of $C'$ sites is
significantly greater than the number of empty $\bar A$ sites. The
situation in glass has some resemblance to the ''vacancy-like'' models proposed
recently by several authors \cite{schroder02,jain02,lammert03}. Under the
influence of external pressure it may even be possible to convert some of
the remaining empty $\bar A$ into smaller $C'$ sites. Elsewhere, one of us
\cite{ingram_bunsen} argues that the apparent ''activation volume'' for
charge transport in cation-conducting glasses can be interpreted in this
way, and indeed is equal to the volume difference between empty $C'$ and
$\bar A$ sites.

Regarding the existence of prefered pathways or conducting
channels \cite{Sunyer03, Meyer04, Kargl04}: In a recent paper, Meyer et al.
\cite{Meyer04} identify a prepeak observed in both neutron scattering data
and molecular dynamics simulations with the emergence of Na$^+$ ion channels
in sodium silicate melts and glasses. In terms of the new DSM, we expect that
$\bar Na$ and $C'$ sites close to the nBOs form these channels.
Hence, suppression of the formation of nBOs, e.g. by addition of $Al_2O_3$ 
to sodium silicate glasses should disrupt the pathway structure, in agreement 
with very recent experiments by Kargl et al. \cite{Kargl04}.

\section{Mixed cation glasses} 
\label{mixed_cat} 
\subsection{The mixed alkali effect (MAE)} 
\label{MAE} 
 
The mixed alkali effect \cite{isard68,day76,imre02} embraces a wide range of
phenomena. Some of these, such as diffusivity crossovers and conductivity
minima are directly related to changes in ion mobility, and are associated
with maxima in activation energies and entropies, and in the activation
volume \cite{Bandaranayake02}. There are, however, other anomalies, which
only indirectly involve ion transport. These include minima in $T_g$
\cite{day76,ingram03}, large internal friction peaks \cite{day76,roling98}
and volumetric relaxations \cite{day76,dietzel83} observed below $T_g$, and
viscosity minima observed in molten glass \cite{poole49}.
 
\unitlength 1.85mm 
\vspace*{0mm} 
{ 
\begin{figure} 
\begin{picture}(80,50) 
\def\epsfsize#1#2{0.26#1} 
\put(5,-1){\epsfbox{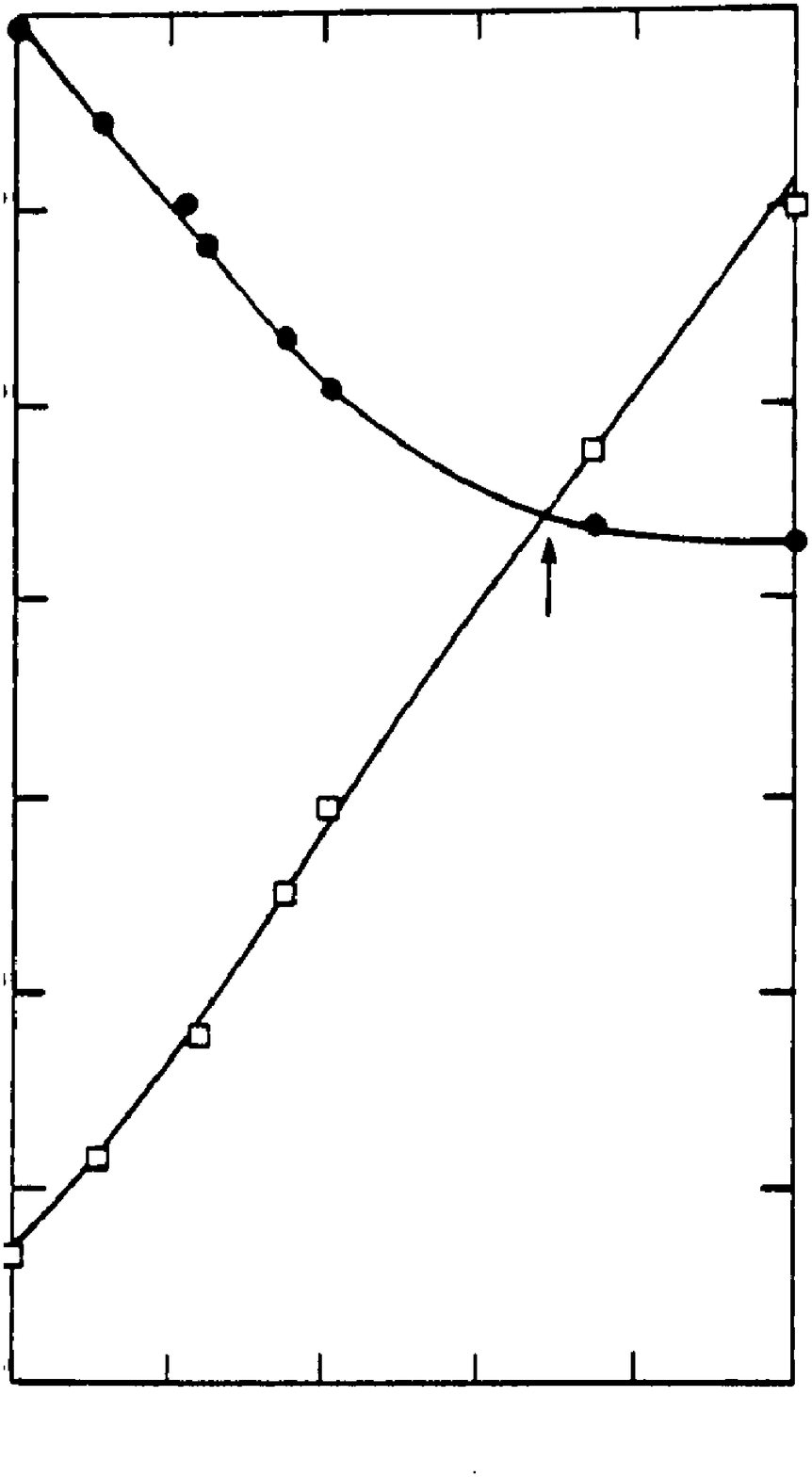}} 
\def\epsfsize#1#2{0.33#1} 
\put(50,2){\epsfbox{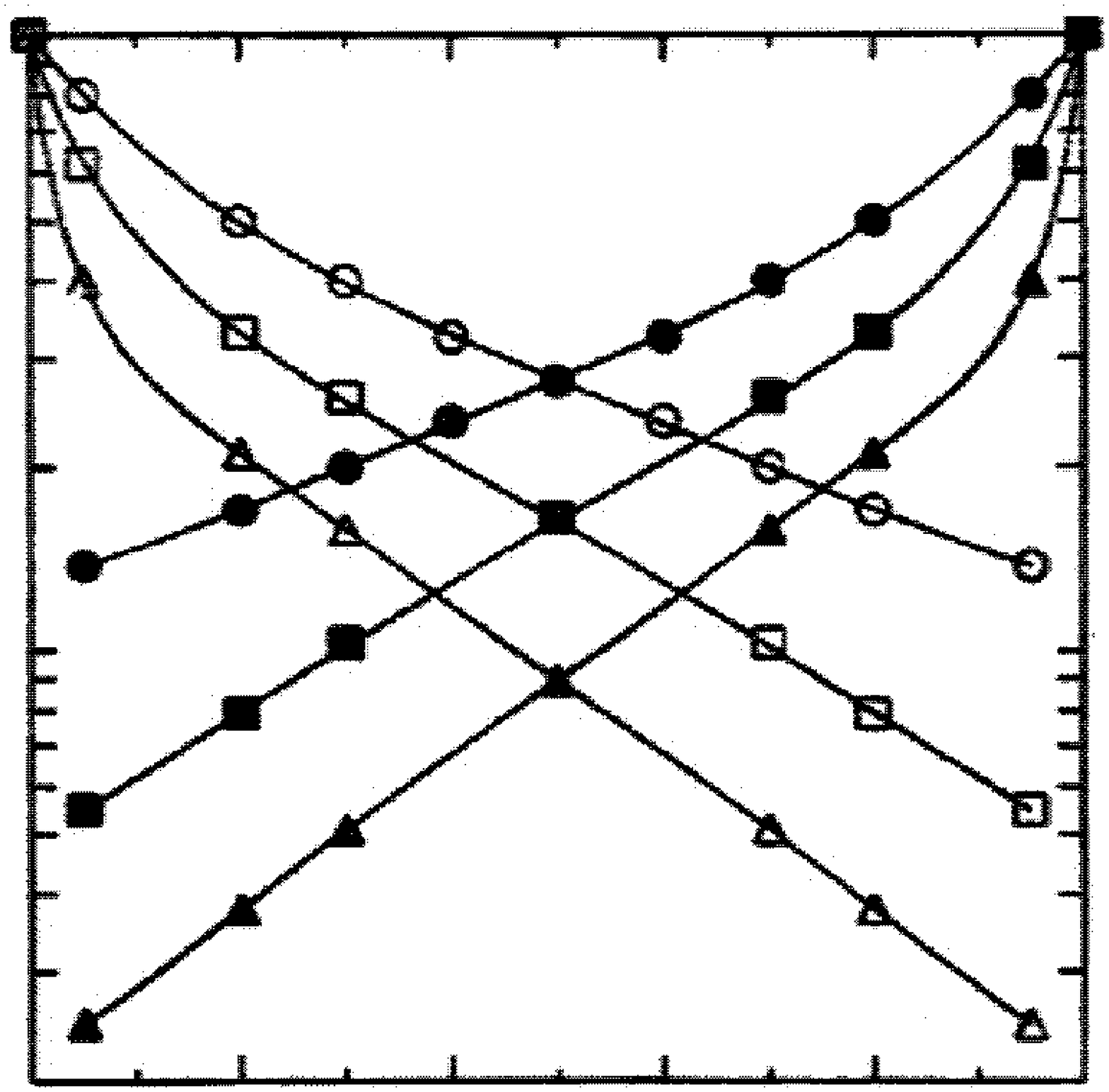}} 
\put(27,9){\makebox(1,1){\bf\large (a)}}  
\put(73,9){\makebox(1,1){\bf\large (b)}} 
\put(25,0){\makebox(1,1){\bf $Cs/(Cs+Na)\quad[\%]$}}  
\put(75,0){\makebox(1,1){\bf $x$}}  
\put(10.5,3){\makebox(1,1){\bf\footnotesize $0$}}  
\put(16,3){\makebox(1,1){\bf\footnotesize $20$}}  
\put(20.5,3){\makebox(1,1){\bf\footnotesize $40$}}  
\put(25.5,3){\makebox(1,1){\bf\footnotesize $60$}}  
\put(30.5,3){\makebox(1,1){\bf\footnotesize $80$}}  
\put(36,3){\makebox(1,1){\bf\footnotesize $100$}} 
\put(7.5,5){\makebox(1,1){\bf\footnotesize $10^{-15}$}}   
\put(7.5,11){\makebox(1,1){\bf\footnotesize $10^{-14}$}}  
\put(7.5,17){\makebox(1,1){\bf\footnotesize $10^{-13}$}}    
\put(7.5,23){\makebox(1,1){\bf\footnotesize $10^{-12}$}}    
\put(7.5,29.5){\makebox(1,1){\bf\footnotesize $10^{-11}$}}    
\put(7.5,36){\makebox(1,1){\bf\footnotesize $10^{-10}$}}    
\put(7.5,42){\makebox(1,1){\bf\footnotesize $10^{-9}$}}    
\put(7.5,48){\makebox(1,1){\bf\footnotesize $10^{-8}$}}    
\put(30,27){\makebox(1,1){\footnotesize Mobility}}  
\put(30,25){\makebox(1,1){\footnotesize crossover}}  
\put(20,42){\makebox(1,1){\bf\footnotesize $\quad^{22}Na$}}  
\put(20,15){\makebox(1,1){\bf\footnotesize $\quad^{137}Cs$}}      
\put(53,18.5){\makebox(1,1){\bf\footnotesize $10^{-1}$}}  
\put(53,38){\makebox(1,1){\bf\footnotesize $10^{0}$}}  
\put(56.5,3){\makebox(1,1){\bf\footnotesize $0$}}    
\put(63,3){\makebox(1,1){\bf\footnotesize $0.2$}}  
\put(69.5,3){\makebox(1,1){\bf\footnotesize $0.4$}}      
\put(76.5,3){\makebox(1,1){\bf\footnotesize $0.6$}}      
\put(83,3){\makebox(1,1){\bf\footnotesize $0.8$}}      
\put(89.5,3){\makebox(1,1){\bf\footnotesize $1$}}          
\put(2,10){\begin{sideways} 
          \bf diffusion coefficients $[cm^2/s]$ 
          \end{sideways} 
         } 
\put(48,5){\begin{sideways} 
          \bf normalized diffusion coefficients 
          \end{sideways}  
          }                                                                  
\end{picture} 
\caption[]{\small (a) Diffusion coefficients in $Na/Cs$ trisilicate 
  glasses at $396\,^\circ C$ as a function of the composition dependence 
  $Cs/(Cs+Na)$. (Data after \cite{jain83}, redrawn from
  Ref.~\cite{ingram87}).  
  (b) Numerical results from the DSM for the normalized diffusion 
  coefficients $D_A(x)/D_A(1)$ and $D_B(x)/D_B(0)$ of cations $A$ and 
  $B$ as a function of $x$ for a symmetric choice of parameters 
  and for three different temperatures. The full symbols refer to 
  the $A$ ions, the empty symbols to the $B$ ions 
  (redrawn after Ref.~\cite{bunde94}).  }
\label{bi:diff} 
\end{figure}} 
  
Figure~\ref{bi:diff}(a) shows the diffusivity crossover in $Na-Cs$ silicate
glasses reported by Jain and coworkers \cite{jain83}. The behaviour of $Na$
differs markedly from that of $Cs$. While the self-diffusion coefficient of
the latter decreases exponentially with decreasing $Cs$ content, the
corresponding coefficient for $Na$ levels off at low $Na$ contents. This
asymmetry is disguised in conductivity plots, which tend to be rather U or
V-shaped in appearance. The conductivity is, however, always dominated by
the more mobile and numerous (majority) cations; the significant differences
in behaviour between large and small ions tend to show up more clearly only
when the ions in question are in the minority.
  
On the other hand it is possible to overstate the degree of asymmetry.
Measurements of activation volumes ($V_A =-RTd\ln\sigma/dP$, with the
pressure $P$ and the gas constant $R$) in mixed cation glasses reveal
increases in $V_A$ occurring at both ends of the composition range, when
larger ions are diluted by smaller ones or vice versa
\cite{Bandaranayake02,voss03}. This evidence for some degree of symmetry has
suggested some degree of coupling between the local motions of the ions,
involving $A^+$ ions moving into $\bar B$ sites and $B^+$ ions moving into
$\bar A$ sites \cite{Bandaranayake02,ingram03}.
 
In the dynamic structure model, the asymmetry was expressed in terms of
different jump rates of $A^+$ and $B^+$ cations. In the published
simulations \cite{bunde94}, for simplicity and for revealing the crucial
effect of mismatch between unlike ions, a hypothetical case was considered
of a glass containing a mixture of cations, $A^+$ and $B^+$, alike in all
respects except that a mismatch energy appeared whenever either ion entered
sites belonging to the other. Figure~\ref{bi:diff}(b) shows the crossover in
diffusivities predicted by the quantitative model \cite{bunde94}. This
figure correctly reveals the steep falls in ion mobility found in the dilute
foreign alkali region, but of course it could not replicate the contrasting
behaviours of $Na$ and $Cs$ shown in Figure~\ref{bi:diff}(a).
 
It is now appropriate to look again at the MAE in the light of the newly
''updated'' dynamic structure model.
 
\subsection{The MAE in molten glass} 
\label{MAE_melt} 
  
There have been several reports in the literature of the MAE found in molten
glass \cite{endell42,baucke89}.  Baucke and Werner have shown, that for a
series of $Na_2O-K_2O-CaO-SiO_2$ glasses, there are significant deviations
from additivity in ionic conductivity, and especially in the activation
energy well above $T_g$ (see Fig.~\ref{bi:cond_acti}(a) and (b),
respectively). The effect is thus still apparent in melts of relatively
modest viscosity, arguably close to the critical temperature, $T_c$, of mode
coupling theory. The persistence of the MAE up to such high temperatures is
strong evidence that distinctions between $\bar A$ and $\bar B$ sites can
still be made.
 
According to our updated dynamic structure model, the sites are formed in
the melt between $T_c$ and $T_g$ where the volumes of the $\bar A$ and the
$C'$ sites do not differ very much. In mixed alkali melts, the situation is
further complicated owing to the different sizes of the $A^+$ and the $B^+$
ions. When the smaller $A^+$ ions are in the overwhelming majority, the $C'$
sites are on average smaller than when the larger $B^+$ ions are in the
majority. Close to the conduction minimum, where nearly the same numbers of
$A^+$ and $B^+$ ions are present, we would expect a broad distribution of
$C'$ sites, including sites to be found in each of the single alkali melts.
Under such conditions, ion transport involves continual melt rearrangement,
and this gives rise to the maximum in activation energy which is so evident
in Baucke and Werner's data.
 
\unitlength 1.85mm 
\vspace*{0mm} 
{ 
\begin{figure} 
\begin{picture}(80,30) 
\def\epsfsize#1#2{0.35#1} 
\put(0,0){\epsfbox{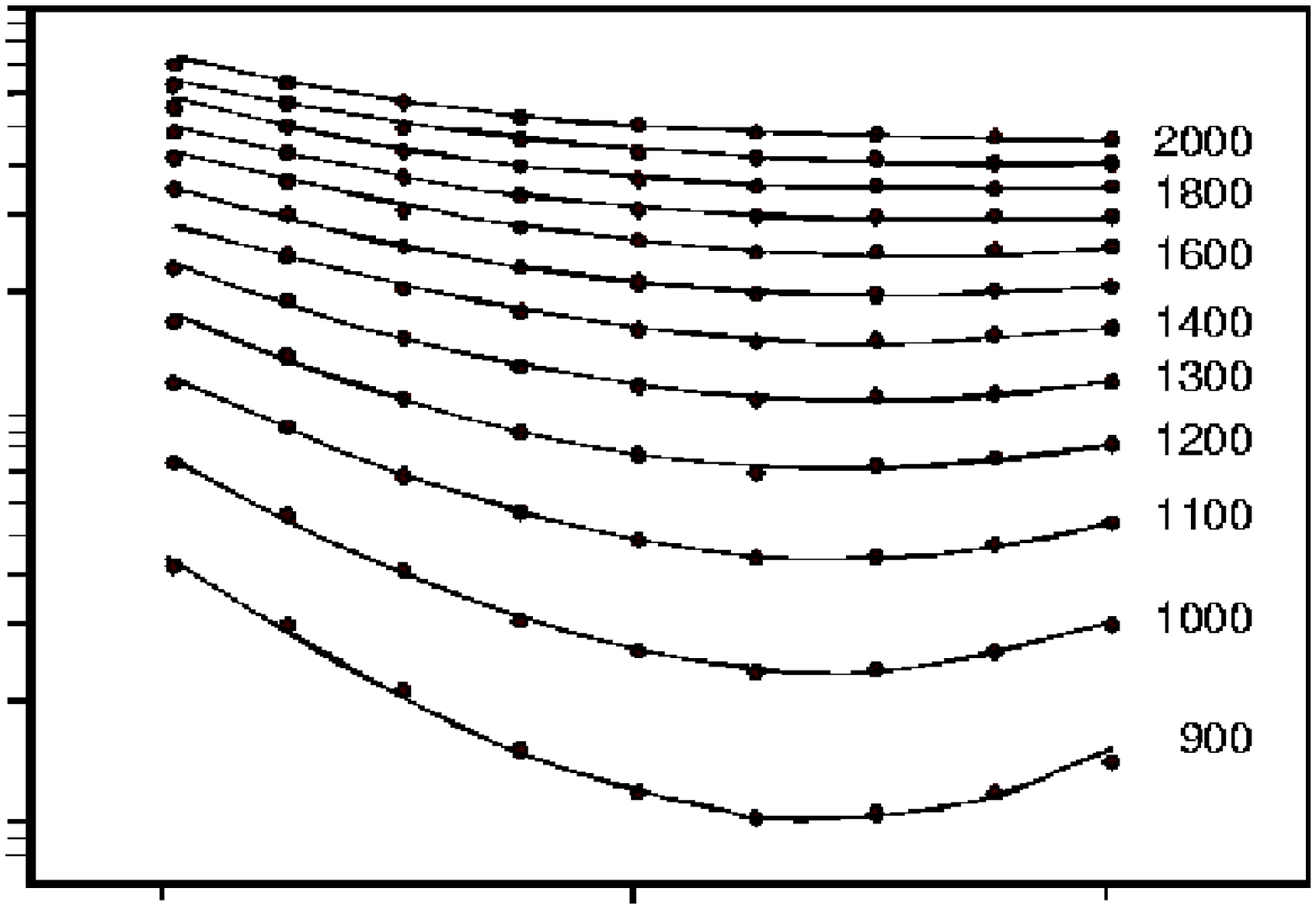}} 
\put(45,0){\epsfbox{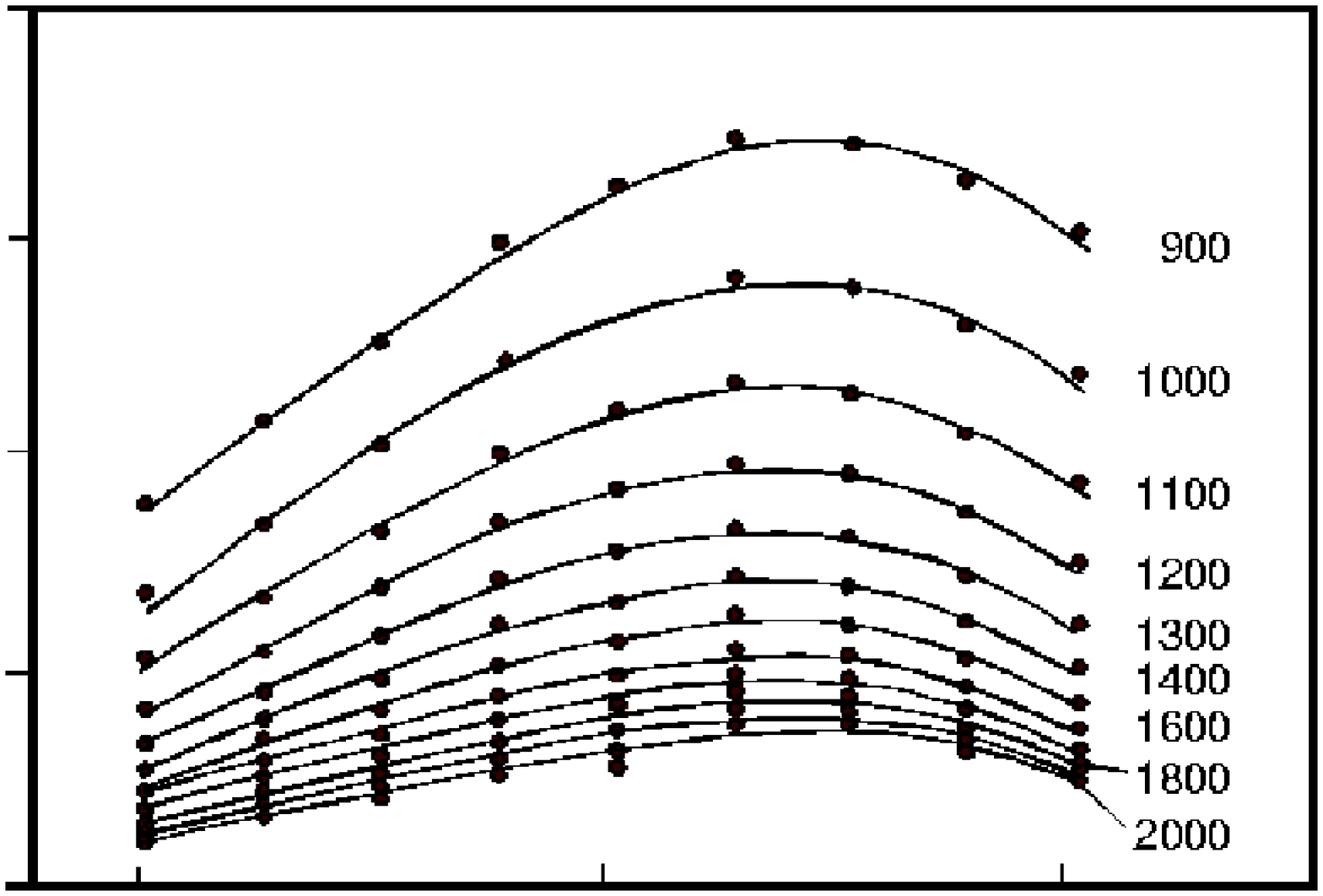}} 
\put(55,26){\makebox(1,1){\bf\large (b)}}  
\put(10,7){\makebox(1,1){\bf\large (a)}}  
\put(30,0){\makebox(1,1){\bf $x$}}  
\put(75,0){\makebox(1,1){\bf $x$}}  
\put(10.5,2){\makebox(1,1){\bf\footnotesize $0$}}  
\put(23.9,2){\makebox(1,1){\bf\footnotesize $0.5$}}  
\put(37.4,2){\makebox(1,1){\bf\footnotesize $1.0$}}  
\put(3.8,5.8){\makebox(1,1){\bf\footnotesize $10^{-2}$}}  
\put(3.8,17.4){\makebox(1,1){\bf\footnotesize $10^{-1}$}}  
\put(3.8,29.2){\makebox(1,1){\bf\footnotesize $10^{0}$}}  
 
\put(55.2,2){\makebox(1,1){\bf\footnotesize $0$}}  
\put(68.5,2){\makebox(1,1){\bf\footnotesize $0.5$}}  
\put(81.5,2){\makebox(1,1){\bf\footnotesize $1.0$}}  
\put(49.7,4.4){\makebox(1,1){\bf\footnotesize $40$}}  
\put(49.7,10.4){\makebox(1,1){\bf\footnotesize $60$}}  
\put(49.7,16.5){\makebox(1,1){\bf\footnotesize $80$}}  
\put(49.7,22.8){\makebox(1,1){\bf\footnotesize $100$}}  
\put(49.7,29.1){\makebox(1,1){\bf\footnotesize $120$}}  

\put(85,27){\makebox(1,1){\bf\footnotesize $T/\,^\circ C$}}  
\put(40,27){\makebox(1,1){\bf\footnotesize $T/\,^\circ C$}}  

\put(0,2){\begin{sideways} 
          \bf Conductance $\sigma$ [$\Omega^{-1}\,cm^{-1}$] 
          \end{sideways} 
         } 
\put(46,3){\begin{sideways} 
          \bf Act. energy [$E_a\,kJ\,mol^{-1}$] 
          \end{sideways}  
          }                                                                 
\end{picture} 
\caption[]{\small (a) Conductivity as a function of the potassium mole fraction $x$  
  in the molten mixed-alkali system $(1-x) Na_2O\cdot xK_2O\cdot 0.7CaO\cdot
  4.8SiO_2$ at various temperatures.  (b) The corresponding activation
  energies at various temperatures taken from curved Arrhenius plots.
  (Redrawn after \cite{baucke89}).  }
\label{bi:cond_acti} 
\end{figure}}

\subsection{The MAE in solid glasses} 
\label{MAE_solid} 
 
The starting point, as in earlier discussions \cite{bunde94}, is to
recognise the mismatch effect which arises whenever ions enter ''wrong
sites''. It is clear that larger ($B^+$) ions will find it harder (as a
natural consequence of electron repulsions) to enter smaller ($\bar A$)
sites than vice versa. Also, because of the broad distribution of $C'$ sites
which formed in the melt, there are many more $C'$ sites available to $A^+$
ions than there are to $B^+$ ions. Accordingly, in $Na/Cs$ glasses (see
Fig.~\ref{bi:diff}(a)), $Cs$ ions will be expected to suffer much larger
losses in mobility than $Na$ ions, which indeed is the experimental result.
This explains also why $Na$ ion mobilities level off at low $Na$ contents,
but to a value lower than that of the majority $Cs$ ions (if ions jump into
sites which are too large for them, they will tend to ''roll back'' into
their original sites.)
 
If cations do not differ too much in size, as in $Li/Na$ and $Na/K$ mixtures
for example, the $A^+$ cations may find themselves entering $\bar B$ sites
in preference to less attractive $C'$ sites. In this way, $A^+$ ions and
$\bar A$ sites become randomly mixed up with $B^+$ ions and $\bar B$ sites
(as indeed is indicated both by infrared and NMR
spectroscopies)\cite{ratai02,kamitsos91}. The frequent entry of $A^+$ ions
into $\bar B$ sites and vice versa leads to a loosening of the structure.
This loosening was seen by Ingram and Roling as being the underlying cause
of minima in $T_g$ \cite{ingram03} and of melt viscosity \cite{poole49}, as
interpreted with the concept of matrix mediated coupling (CMMC).
 
Remarkably, these indirect effects may be smaller when cations differ more
markedly from each other. This was found in internal friction experiments
\cite{day76}. Very small loss peaks were reported in $Li/Cs$ silicate
glasses, compared with much larger ones in $Li/Na$ glasses. We now
understand this additional anomaly (i.e. the suppression of an important
aspect of the mixed alkali effect) in terms of $Li^+$ and $Cs^+$ ion simply
differing too much from each other to enter each other's sites.
 
\section{Conclusions} 
\label{concl} 
 
We introduce an updated dynamic structure model for explaining the complex
behaviour of single and mixed cation glasses. As in the original version
\cite{maass92,bunde94}, we distinguish between $\bar A$, $\bar B$ and $\bar
C$ sites. However, in contrast to the old DSM, we assume that $\bar A$ and
$\bar B$ sites can only be generated close to the countercharges (which
often are non-bridging oxygens). When an $\bar A$ site is vacated, it
relaxes to a $C'$ site whose size is determined largely by the host matrix,
but it is better suited to accommodate mobile ions than are the $\bar C$
sites, which are remote from the localized countercharges. In a single
cation glass, the $\bar A$ and $C'$ sites found close to the countercharges,
form ''stepping stones'' for the mobile $A^+$ ions.

As in the recent papers \cite{Sunyer03, Meyer04, Kargl04}, we expect
that these pathways will appear in the melt, perhaps at a
temperature close to $T_c$ of mode coupling theory, when the diffusion of
the network forming units is blocked, and viscosity becomes very temperature
dependent. These pathways constitute an integral part of glass structure
below $T_g$. The pathways become interrupted when the stepping stones are
too far apart, i.e. when the contercharge concentrations 
are too small. The pathways may be ''frozen in'' at $T_g$ or they may
be allowed to fluctuate in the glassy state as a result of localised density
fluctuations. In both cases, the loss of good pathways leads to the abnormal 
decreases in
conductivity with decreasing ion content, an effect which is especially
severe in certain systems such as alkali borate glasses.
  
In a mixed ionic glass, it is much more difficult to find separate pathways
for the $A^+$ and the $B^+$ ions, made up of $\bar A$ and $C'$ and of $\bar
B$ and $C'$ sites, respectively. This appears to be the underlying cause of
many of the observed anomalies. As in the old DSM, we envisage that there
will be a mismatch whenever a (small) $A^+$ ion enters a (large) $\bar B$
site or vice versa. We now recognise that this mismatch is greater if a
larger cation is trying to enter a smaller site, and in this way we can
account for the observed asymmetries in diffusion behaviour, these being
especially significant with cations of markedly different sizes, as in the
$Na-Cs$ system for example. By contrast, the smaller of the two ions ($A^+$)
will make use of available $C'$ sites, when the supply of ($\bar A$) sites
is insufficient to provide extended diffusion pathways. This comment is
consistent with the ''levelling off'' in the diffusion coefficient of $Na^+$
ions at low concentrations, as seen in the $Na-Cs$ system.
 
However, in molten glass, the $A^+$ and $B^+$ ions will have sufficient
energy to visit each other's sites, and regularly disturb the local
structure. This could possibly explain the strong reduction of $T_g$ and of
melt viscosity reported in many mixed alkali systems. When the ions do not
differ too much in size, these site exchange effects may persist below
$T_g$. In this way, we could account for the large internal friction peaks
seen in $Li/Na$ silicate glasses and for the anomalous increases in
activation volume which point to some coupling between the motions of the
$A^+$ and the $B^+$ cations.
 
What is urgently needed is more information about empty sites in glass, and
how the nature of these sites depends on glass composition and temperature.
In the related field of polymer electrolytes, diagnostic information is
coming from positron annihilation lifetime spectroscopy or PALS
\cite{forsyth00,bamford03}. This technique exploits the variable lifetimes
of the ortho-positronium species in order to determine the size of available
cavities within the host matrix. Such information, together with activation
volumes determined from variable pressure measurements, might enable $\bar
A$ and $\bar B$ sites to be distinguished from each other and from the
possibly more numerous $C'$ sites.  Comparisons made between fast quenched
and slow cooled glasses would provide information not only on structural
dynamics but also on the growth and consolidation of percolation pathways.

Finally, it is interesting to note previous comments in the literature to
the effect that the concentration dependence of conductivity in mixed cation
glasses is closely related to those encountered in single cation systems
\cite{sakka79}. If this rule applies quite generally, it would imply that
the $C'$ sites encountered in single cation glasses must be very similar to
those encountered in mixed cation systems. This may or may not be the
case. Further consideration of the MAE in such glass forming systems as the
aluminosilicates \cite{lapp87,ingram98}, where the conductivity dependence
in single cation glasses is weak \cite{Hunter00}, should be highly
informative.
 
\bigskip 
 
Acknowledgements: 
 
We gratefully acknowledge financial support to MDI from the Humboldt
Foundation and valuable discussions with Philipp Maass, Andreas Heuer,
Cornelia Cramer and other members of the SFB 450 at the university
of M\"unster.

  
  

\end{document}